# Wind from the black-hole accretion disk driving a molecular outflow in an active galaxy

F. Tombesi[1,2], M. Melendez[2], S. Veilleux[2,3], J. N. Reeves[4,5], E. González-Alfonso[6], C. S. Reynolds[2,3]

[1]X-ray Astrophysics Laboratory, NASA/Goddard Space Flight Center, Greenbelt, MD, 20771, USA

[2]Department of Astronomy and CRESST, University of Maryland, College Park, MD, 20742, USA

[3]Joint Space Science Institute, University of Maryland, College Park, MD20742, USA

[4]Astrophysics Group, School of Physical and Geographical Sciences, Keele University, Keele, Staffordshire ST5 5BG, UK

[5]Center for Space Science and Technology, University of Maryland Baltimore County, 1000 Hilltop Circle, Baltimore, MD 21250, USA

[6]Universidad de Alcalá, Departamento de Física y Matemáticas, Campus Universitario, E-28871 Alcalá de Henares, Madrid, Spain

**Powerful winds driven by active galactic nuclei (AGN) are often invoked to play a fundamental role in the evolution of both supermassive black holes (SMBHs) and their host galaxies, quenching star formation and explaining the tight SMBH-galaxy relations[1,2]. Recent observations of large-scale molecular outflows[3,4,5,6,7,8] in ultraluminous infrared galaxies (ULIRGs) have provided the evidence to support**

**these studies, as they directly trace the gas out of which stars form. Theoretical models[9,10,11,12] suggest an origin of these outflows as energy-conserving flows driven by fast AGN accretion disk winds. Previous claims of a connection between large-scale molecular outflows and AGN activity in ULIRGs were incomplete[3,4,5,6,7,8] because they were lacking the detection of the putative inner wind. Conversely, studies of powerful AGN accretion disk winds to date have focused only on X-ray observations of local Seyferts[13,14] and a few higher redshift quasars[15,16,17,18,19]. Here we show the clear detection of a powerful AGN accretion disk wind with a mildly relativistic velocity of 0.25$c$ in the X-ray spectrum of IRAS F11119+3257, a nearby (z = 0.189) optically classified type 1 ULIRG hosting a powerful molecular outflow[6]. The AGN is responsible for ~80% of the emission, with a quasar-like luminosity[6] of $L_{AGN} = 1.5 \times 10^{46}$ erg s$^{-1}$. The energetics of these winds are consistent with the energy-conserving mechanism[9,10,11,12], which is the basis of the quasar mode feedback[1] in AGN lacking powerful radio jets.**

The mass[20] of the central SMBH in IRAS F11119+3257 is estimated to be $M_{BH}$ ~ 1.6 × 10$^7$ M$_\odot$. The resultant Eddington ratio is $L_{AGN}/L_{Edd}$ ~ 5, indicating that the source is very likely accreting at about its Eddington limit. In the X-rays it is relatively bright and not strongly affected by neutral absorption[21]. Therefore, it is arguably the best candidate to study the highly ionized iron K-band absorbers in this class of object.
IRAS F11119+3257 was the subject of a long 250 ks *Suzaku* observation obtained by our group in May 2013. The spectra of the three separated XIS detectors onboard *Suzaku* independently show a prominent absorption feature at the rest-frame energy of ~9 keV

(Figure 1). The inclusion of a broad Gaussian absorption line in the combined broad-band *Suzaku* XIS and PIN spectrum in the 0.5−25 keV energy band provides a very significant improvement of the fit, corresponding to a statistical confidence level of 6.5σ (see SI). The absorption line has a rest-frame energy of $E = 9.82^{+0.64}_{-0.34}$ keV, a width of $\sigma_E = 1.67^{+1.00}_{-0.44}$ keV and an equivalent width of $EW = -1.31^{+0.40}_{-0.31}$ keV. If interpreted as blue-shifted Fe XXV Heα and/or Fe XXVI Lyα resonant absorption lines, this feature indicates an outflow velocity of ~0.3$c$ with respect to the systemic velocity.

We derive a physical characterization of this outflow using a dedicated photoionization absorption model (see Methods). This fast wind model provides a very good representation of the data, simultaneously taking into account both the broad absorption trough at $E \sim 7-10$ keV in the XIS data and the factor of ~2−3 excess of flux at $E = 15-25$ keV observed in the PIN data (see Extended Data Figure 1 and Methods). The column density, ionization and outflow velocity are $N_H = (6.4^{+0.8}_{-1.3}) \times 10^{24}$ cm$^{-2}$, $\log\xi = 4.11^{+0.09}_{-0.04}$ erg s$^{-1}$ cm and $v_{out,X} = 0.255 \pm 0.011\ c$, respectively (see Extended Data Table 1). The wind has a high covering fraction of $C_{F,X} > 0.85$, confirmed if including the possible associated emission (see SI). The wind parameters are consistent overall with the mildly relativistic accretion disk winds detected in quasars[15,16,17,18,19] with a comparable luminosity.

Considering the velocity and variability of the wind we estimate a distance from the central SMBH in units of Schwarzschild radii ($r_s = 2\ GM_{BH}/c^2$) of $r \approx 15-900\ r_s$, consistent with accretion disk scales (see SI). Conservatively adopting the radius of 15 $r_s$, we estimate a mass outflow rate of $\dot{M}_{out,X} \approx 1.5\ M_\odot$ yr$^{-1}$. Given the very high luminosity of this AGN and the fact that it is accreting at about its Eddington limit, it is very likely that the wind is radiation driven[22].

The wind's momentum flux $\dot{P}_{out,X} = \dot{M}_{out,X} v_{out,X} \approx 6 \times 10^{35}$ dyne corresponds to $1.3^{+1.7}_{-0.9}$ times that of the AGN radiation $\dot{P}_{rad} = L_{AGN}/c$ (see SI). This is consistent with multiple scatterings in a high column and wide angle wind[23]. The wind mechanical power $\dot{E}_{K,X} = (1/2)\dot{M}_{out,X} v_{out,X}^2 \approx 2 \times 10^{45}$ erg s$^{-1}$ corresponds to ~15% of the AGN luminosity and it is higher than the minimum value of ~0.5% required for AGN feedback[24] (see SI). Therefore this wind will likely have a strong influence on the host galaxy environment.

An alternative interpretation of the E ~ 9 keV absorption trough with an ionized Fe K edge and an associated relatively slow wind (see Extended Data Figure 1) is statistically disfavored (see Methods).

Likewise the case whereby the Fe K profile arises from relativistically blurred X-ray reflection off the inner accretion disk is also ruled out (see Methods and SI). Such a model fails to predict the hard X-ray emission above 10 keV (Extended Data Figure 1). This case is also inconsistent with the variability observed during the observation (see Methods and Extended Data Figure 2 and 3). The fast wind model provides a self-consistent explanation of the variability in terms of a change in the absorber column density (Extended Data Table 2). In contrast, within the relativistic reflection model, the resultant variability pattern is the opposite of that expected from relativistic light bending[25], with a dramatic increase in reflection fraction from the low to the high flux state (see Extended Data Table 3). Finally, the extrapolated hard X-ray luminosity from the fast wind model matches the value expected from mid-infrared emission line diagnostics, while the relativistic reflection model underestimates this value by more than one order of magnitude (see SI).

*Herschel* observations show a massive molecular outflow[6] in the absorption profile of the 119 µm OH doublet of IRAS F11119+3257 (Figure 2). This galaxy also shows clear signatures of high velocity outflows (~1,000 km s$^{-1}$) at other energies, including neutral[26] and moderately ionized[27,28] gas. From our radiative transfer model[7] of the *Herschel* observation of the OH 119 absorption profile (see Methods and Figure 2) we estimate a 300 pc scale outflow with average velocity of $v_{out,OH}$ = 1,000 ± 200 km s$^{-1}$, covering factor $C_{F,OH}$ = 0.20 ± 0.05 and mass outflow rate of $\dot{M}_{out,OH} = 800^{+1,200}_{-550}$ M$_\odot$ yr$^{-1}$ (see SI). The momentum flux of the molecular outflow is considerably larger than that of the AGN radiation, $\dot{P}_{out,OH} = 11.0^{+14.1}_{-7.5} \dot{P}_{rad}$. Moreover, the mechanical power corresponds to ~2% of the AGN luminosity (see SI), a value higher than the minimum required for quasar mode feedback[24]. The large velocity and energetics of this molecular outflow also favor an AGN origin[3,4,5]. These parameters are comparable to those found in similar AGN dominated ULIRGs[3,4,5,6,7,8].

The two main mechanisms invoked to explain AGN feedback[1,2] are the radio mode and the quasar or wind mode. In these cases the molecular outflows are accelerated by the interaction with either a powerful, but highly collimated, radio jet[29] or a mildly relativistic accretion disk wind[9,10,11,12], respectively. Powerful jets are found in radio-loud AGNs, but they constitute only a minority of the AGN population[29]. IRAS F11119+3257, and ULIRGs in general, do not show clear evidence for powerful jets[6,8,30]. Therefore, the radio mode feedback is strongly disfavored in this case.

Theoretical models of quasar mode AGN feedback[9,10,11,12,22] indicate that the shock caused by the interaction between the putative mildly relativistic AGN wind and the interstellar medium divides the resultant large-scale outflow into two regimes. Momentum-conserving flows occur if the shocked wind gas can cool and most of the kinetic energy is radiated away. On the other hand, energy-conserving flows occur if the shocked wind gas is not efficiently cooled, and instead expands as a hot bubble. Conservation of energy leads to the relation[9,10] $\dot{P}_{out} = f\,(v_{in}/v_{out})(L_{AGN}/c)$ where $v_{in}$ is the velocity of the inner X-ray wind, $v_{out}$ the velocity of the molecular outflow and $f = C_{F,out}/C_{F,in}$ is the ratio between the covering fractions of the outer molecular outflow and inner disk wind. The latter parameter indicates the efficiency or fraction of inner wind power that is transferred to the large-scale outflow. Substituting the values from the observed X-ray and OH winds in IRAS F11119+3257 we derive an efficiency of $f = 0.22 \pm 0.07$.

In Figure 3 we show the momentum fluxes of the inner X-ray wind and the molecular outflow observed in IRAS F11119+3257 with respect to their velocities, normalized to the momentum flux of the AGN radiation. For comparison, we consider also the energetics of mildly relativistic X-ray winds detected in quasars[15,16,17,18,19] and the molecular outflows detected in ULIRGs[3,4,5,6,7,8] with similar AGN luminosities (see SI and Extended Data Table 4). We consider the derived efficiency of $f = 0.2$. Using the relation for an energy-conserving flow[9,10,11] we find that the energetics of the molecular outflows would require initial AGN winds with a range of velocities between $v_{in} \sim$ 0.1−0.4$c$ (Figure 3). Indeed, substituting the estimated values for IRAS F11119+3257 of

$v_{in} \approx 0.2c$, $v_{out} \approx 1{,}000$ km s$^{-1}$ and $f \approx 0.2$ we obtain a value of $\dot{P}_{out} \approx 12$ $L_{AGN}/c$ consistent with the OH observation.

This would not be possible in the momentum-conserving case. In fact, considering that a fraction $f = 0.2$ of the initial wind momentum rate is transferred to the host and taking into account the uncertainties, the maximum value for the inner disk wind is $f\dot{P}_{out,X} = 0.6$ $L_{AGN}/c$. This is much smaller than the minimum value for the molecular outflow of $\dot{P}_{out,OH} = 3.5$ $L_{AGN}/c$ (see SI and Figure 3).

The mass outflow rate of radiation driven winds[9,22] is $\dot{M}_{in} \sim \dot{M}_{Edd}$, where $\dot{M}_{Edd} = L_{Edd}/\eta c^2$ is the Eddington accretion rate and $\eta$ is the accretion efficiency. Moreover, their momentum rate is comparable[9,22,23] to that of the AGN radiation, $\dot{P}_{in} \sim \dot{M}_{in} v_{in} \sim L_{Edd}/c$. Therefore, the observed fast wind velocity suggests[9] a high accretion efficiency $\eta \sim v_{in}/c \sim 0.25$. For a stellar velocity dispersion of $\sigma \sim 100$ km s$^{-1}$ from the $M_{BH}$–$\sigma$ relation[9,11], theoretical models of energy-conserving outflows[9] suggest a molecular outflow velocity of $v_{out} \sim 1000$ km s$^{-1}$, mass outflow rate of $\dot{M}_{out} \sim 1000$ $M_{\odot}$ yr$^{-1}$ and a momentum rate $\dot{P}_{out} \sim 10$–$20$ $\dot{P}_{rad}$, in agreement with the observations. Moreover, the mechanical energy of the fast wind of ~15% $L_{AGN}$ is consistent with the theoretical value[9] of $(\eta/2)L_{AGN}$ ~13% $L_{AGN}$. The observed power of the molecular outflow of ~2% $L_{AGN}$ is also consistent with that of the fast wind with an efficiency $f \sim 0.2$. With the limitation that the fast X-ray wind is observed now, while the large-scale molecular outflow is probably an integrated effect of such winds over a much longer period of time, there is a very good quantitative agreement between observations and theoretical models[9,10,11,22].

This supports the idea that AGN winds can indeed provide an efficient way to transfer energy to the interstellar medium with a high degree of isotropy[12], as required by the existence of large-scale outflows in most ULIRGs[3,4,5,6,7,8]. Moreover, these results are consistent with the ULIRG evolutionary scheme[6] where AGN driven winds can clear out the obscuring material from the central regions of the galaxy[1,2], eventually uncovering the underlying quasar.

**References**

1. Fabian, A. Observational evidence of active galactic nuclei feedback. *Ann.Rev.Astron.Astrophys*. **50**, 455 – 489 (2012)

2. Veilleux, S., Cecil, G., Bland-Hawthorn, J. Galactic Winds. *Ann.Rev.Astron.Astrophys*. **43**, 769 – 826 (2005)

3. Fischer, J. *et al.* Herschel-PACS spectroscopic diagnostics of local ULIRGs: Conditions and kinematics in Markarian 231. *Astron. Astrophys.* **518,** L41 (2010)

4. Feruglio, C., Maiolino, R., Piconcelli, E., Menci, N., Aussel, H. *et al.* Quasar feedback revealed by giant molecular outflows. *Astron. Astrophys.* **518,** LL155 (2010)

5. Sturm, E. *et al.* Massive Molecular Outflows and Negative Feedback in ULIRGs Observed by Herschel-PACS. *Astrophys. J.* **733**, L16 (2011)

6. Veilleux, S. *et al.* Fast Molecular Outflows in Luminous Galaxy Mergers: Evidence for Quasar Feedback from Herschel. *Astrophys. J.* **776**, 27 (2013)

7. González-Alfonso, E. *et al.* The Mrk 231 molecular outflow as seen in OH. *Astron. Astrophys.* **561,** A27 (2014)

8. Cicone, C. *et al.* Massive molecular outflows and evidence for AGN feedback from CO observations. *Astron. Astrophys.* **562,** A21 (2014)

9. Zubovas, K., King, A. Clearing Out a Galaxy. *Astrophys. J.* **745**, L34 (2012)

10. Faucher-Giguère, C.-A., Quataert, E. The physics of galactic winds driven by active galactic nuclei. *Mon. Not. R. Astron. Soc.* **425,** 605-622 (2012)

11. Zubovas, K., Nayakshin, S. Energy- and momentum-conserving AGN feedback outflows. *Mon. Not. R. Astron. Soc.* **440,** 2625-2635 (2014)

12. Wagner, A. Y., Umemura, M., Bicknell, G. V. Ultrafast Outflows: Galaxy-scale Active Galactic Nucleus Feedback. *Astrophys. J.* **763**, L18 (2013)

13. Tombesi, F. *et al.* Evidence for ultra-fast outflows in radio-quiet AGNs. I. Detection and statistical incidence of Fe K-shell absorption lines. *Astron. Astrophys.* **521,** A57 (2010)

14. Gofford, J. *et al.* The Suzaku view of highly ionized outflows in AGN - I. Statistical detection and global absorber properties. *Mon. Not. R. Astron. Soc.* **430,** 60–80 (2013)

15. Chartas, G., Saez, C., Brandt, W. N., Giustini, M., Garmire, G. P. Confirmation of and Variable Energy Injection by a Near-Relativistic Outflow in APM 08279+5255. *Astrophys. J.* **706**, 644−656 (2009)

16. Pounds, K. A., Reeves, J. N. Quantifying the fast outflow in the luminous Seyfert galaxy PG1211+143. *Mon. Not. R. Astron. Soc.* **397,** 249−257 (2009)

17. Lanzuisi, G. *et al.* HS 1700+6416: the first high-redshift unlensed narrow absorption line-QSO showing variable high-velocity outflows. *Astron. Astrophys.* **544,** A2 (2012)

18. Chartas, G. *et al.* Magnified Views of the Ultrafast Outflow of the $z = 1.51$ Active Galactic Nucleus HS 0810+2554. *Astrophys. J.* **783**, 57 (2014)

19. Gofford, J. *et al.* Revealing the Location and Structure of the Accretion Disk Wind in PDS 456. *Astrophys. J.* **784**, 77 (2014)

20. Kawakatu, N., Imanishi, M., Nagao, T. Anticorrelation between the Mass of a Supermassive Black Hole and the Mass Accretion Rate in Type 1 Ultraluminous Infrared Galaxies and Nearby QSOs. *Astrophys. J.* **661**, 660−671 (2007)

21. Teng, S. H., Veilleux, S. X-QUEST: A Comprehensive X-ray Study of Local ULIRGS and QSOs. *Astrophys. J.* **725**, 1848-1876 (2010)

22. King, A. R., Pounds, K. A. Black hole winds. *Mon. Not. R. Astron. Soc.* **345,** 657-659 (2003)

23. Reynolds, C. S. Constraints on Compton-thick Winds from Black Hole Accretion Disks: Can We See the Inner Disk? *Astrophys. J.* **759**, L15 (2012)

24. Hopkins, P. F., Elvis, M. Quasar feedback: more bang for your buck. *Mon. Not. R. Astron. Soc.* **401,** 7-14 (2010)

25. Miniutti, G., Fabian, A. C. A light bending model for the X-ray temporal and spectral properties of accreting black holes. *Mon. Not. R. Astron. Soc.* **349,** 1435-1448 (2004)

26. Rupke, D. S., Veilleux, S., Sanders, D. B. Outflows in Active Galactic Nucleus/Starburst-Composite Ultraluminous Infrared Galaxies. *Astrophys. J.* **632**, 751−780 (2005)

27. Lipari, S. *et al.* Extreme galactic wind and Wolf-Rayet features in infrared mergers and infrared quasi-stellar objects. *Mon. Not. R. Astron. Soc.* **340,** 289-303 (2003)

28. Spoon, H. W. W., Holt, J. Discovery of Strongly Blueshifted Mid-Infrared [Ne

III] and [Ne V] Emission in ULIRGs. *Astrophys. J.* **702**, L42−L46 (2009)

29. Tadhunter, C., Morganti, R., Rose, M., Oonk, J. B. R., Oosterloo, T. Jet acceleration of the fast molecular outflows in the Seyfert galaxy IC 5063. *Nature* **511**, 440−443 (2014)

30. Nagar, N. M., Wilson, A. S., Falcke, H., Veilleux, S., Maiolino, R. The AGN content of ultraluminous IR galaxies: High resolution VLA imaging of the IRAS 1 Jy ULIRG sample. *Astron. Astrophys.* **409,** 15−121 (2003)

**Supplementary Information** is linked to the online version of the paper at www.nature.com/nature.

**Acknowledgements** F.T. would like to thank T. Kallman, J. García, F. Tazaki, F. Paerels and M. Cappi for comments. FT acknowledges support from NASA (grant NNX12AH40G). M.M. and S.V. are supported in part by NASA grants NHSC/JPL RSA 1427277 and 1454738. S.V. also acknowledges partial support through grant NSF-AST1009583. J.N.R acknowledges the financial support of the STFC. E.G-A is a Research Associate at the Harvard-Smithsonian Center for Astrophysics, and thanks the Spanish Ministerio de Economía y Competitividad for support under projects AYA2010-21697-C05-0 and FIS2012-39162-C06-01. CSR thanks support from NASA (grant NNX14AF86G) and the U.S. National Science Foundation (grant AST1333514).

**Author Contributions** F.T. is the Principal Investigator of the *Suzaku* observation. He led the X-ray spectral analysis, interpretation of the results and manuscript preparation. M.M. and E. G.-A. performed the analysis and modeling of the *Herschel* data. S.V. contributed to the interpretation of the results. J.N.R. and C.S.R. contributed to the X-ray spectral analysis and interpretation of the results. All authors participated in the review of the manuscript.

**Author Information** Reprints and permissions information is available at www.nature.com/reprints. The authors declare no competing financial interests. Correspondence and requests for materials should be addressed to F.T. (ftombesi@astro.umd.edu)

**Figure 1. Absorption line in the *Suzaku* spectrum of IRAS F11119+3257.** Ratio between the separated *Suzaku* XIS0 (black), XIS1 (red) and XIS3 (blue) spectra and the absorbed power-law continuum model in the E=4−12 keV energy range. The ratio of the model including the absorption line at the rest-frame energy of 9.82 keV is superimposed in green. The data are binned to a signal to noise ratio of 10σ for clarity. Errors are at the 1σ level.

**Figure 2. *Herschel*-PACS OH119 observation of IRAS F11119+3257.** The blue (high velocity) and red (systemic velocity) dashed lines represent the different components of the outflow model. The green line represents the total best-fit model. The dashed vertical lines indicate the position of the OH119 doublet components relative to the blue line of the doublet (119.23μm) and with respect to the systemic redshift of the galaxy. The average outflow velocity estimated from the OH model is 1,000 ± 200 km s$^{-1}$.

**Figure 3. Comparison between the inner winds and the molecular outflows.** The momentum flux (dP/dt) normalized to the radiation ($L_{AGN}$/c) is plotted against the wind velocity. The disk and molecular winds in IRAS F11119+3257 (red filled stars), the disk winds of other quasars[15,16,17,18,19] (blue filled circles) and the molecular outflows of other

ULIRGs (OH[5,7] green and CO[4,8] black filled triangles) are reported. Uncertainties are at the 1σ. The black curves represent the energy-conserving trends[9,10,11] and the horizontal gray line indicates the momentum-conserving case. The efficiency $f = 0.2$ is assumed. The momentum rates of the disk winds are multiplied by $f$.

**Methods**

***Suzaku* observation log.** IRAS F11119+3257 was observed with *Suzaku* between 13−19 May 2013 for a total exposure of 250 ks. The data reduction and analysis were performed following the standard procedures as described in the *Suzaku* data reduction guide. We use the heasoft v.6.12 package and the latest calibration files.

***Suzaku* XIS data reduction.** We derived the *Suzaku* XIS cleaned event files and applied standard screening criteria. The 3×3 and 5×5 editing modes were combined. The source spectra were extracted from circular regions of 2.5'' radius centered on the source. The background spectra were extracted from annular regions with inner and outer radii of 3−4'' centered on the source and excluding contamination from the calibration sources. The spectra from the two front illuminated detectors, XIS0 and XIS3, were combined after verifying that the data are consistent with each other. Hereafter we refer to them as XIS03. The data of the back illuminated XIS1 detector are used as well. The XIS response and ancillary files were produced. The source countrates in units of counts s$^{-1}$ in the E = 0.5−10 keV band are 0.0576, 0.0649 and 0.06 for the XIS0, XIS1 and XIS3, respectively. The background countrates are 18% and 30% of the XIS03 and XIS1, respectively.

***Suzaku* PIN data reduction.** The *Suzaku* PIN source spectrum was extracted within the good time intervals and it was corrected for the detector dead time. The latest response file was used. We use the latest and most accurate tuned non X-ray background (NXB) event file version 2.2ver1403 provided by the *Suzaku* team. The NXB event file was used as input to the hxdpinxbpi task and combined with a model of the cosmic X-ray background (CXB) provided by the *Suzaku* team. We conservatively include a systematic uncertainty on the PIN background of 1.5%. The source is detected in the PIN at the 4$\sigma$ level. The source countrate in the E = 15−25 keV band is $0.005 \pm 0.001$ counts s$^{-1}$, including the systematic error on the background. The average source countrate is 3.3% of the background. There are no other X-ray sources with 2−10 keV flux higher than $5 \times 10^{-14}$ erg s$^{-1}$ cm$^{-2}$ in the Chandra, XMM-Newton, BeppoSAX and ASCA archives, excluding the contamination from other sources within the PIN field of view. Above 10 keV, we do not find any source detected within a 1° radius in the Swift BAT 70-month, RXTE and Integral surveys, being also the PIN flux of IRAS F1111+3257 below the sensitivity limit of these hard X-ray surveys.

***Suzaku* spectral analysis.** The spectral analysis was carried out using the software XSPEC v.12.7.1. All uncertainties quoted are at 1$\sigma$ level for one parameter of interest and the energies are reported in the source rest-frame, unless otherwise stated. We exploit the broad-band capabilities of *Suzaku* combining the E=0.5−10.5 keV XIS and E=15−25 keV PIN spectra. We perform joint fits of the *Suzaku* XIS03 and XIS1 spectra excluding the energy range around the Si K edge (E=1.5−2 keV), which is known to be affected by calibration issues. The XIS03/XIS1 cross-normalization was left free to vary, but it was always found to be consistent within 3%. We take into account the XIS/PIN cross-normalization of $1.16 \pm 0.01$. All spectra were grouped to a minimum of 25 counts per energy bin in order to allow the use of the $\chi^2$ minimization in the model fitting. Throughout we include a neutral Galactic absorption of $N_H = 2.1 \times 10^{20}$ cm$^{-2}$.

**Fast wind model.** We model the broad absorption at E ~ 9 keV with the XSTAR[31] code v. 2.2.1bn. We consider a $\Gamma = 2$ power-law continuum, consistent with the observed value (see SI), and standard Solar abundances. A turbulent velocity of 30,000 km s$^{-1}$ is assumed for the fast wind model. This high value is introduced only to model the large width of the absorption line and it is probably not linked to an actual physical turbulence in the gas. For instance, detailed accretion disk wind models show that the line profiles can become significantly broadened because of the velocity shear between consecutive zones of the wind[32].

This provides a very good fit to the data, with $\chi^2/\nu = 1410.4/1391$, providing a higher fit improvement comparable to the phenomenological best-fit case of a broad absorption line ($\Delta\chi^2/\Delta\nu = 54.2/3$). This corresponds to a very high detection confidence level of 6.5σ. The relative column density, ionization and outflow velocity are $N_H = (6.4^{+0.8}_{-1.3}) \times 10^{24}$ cm$^{-2}$, $\log\xi = 4.11^{+0.09}_{-0.04}$ erg s$^{-1}$ cm and $v_{out,X} = 0.255 \pm 0.011\ c$, respectively (see Extended Data Table 1). The absorber is consistent with fully covering the source, with a lower limit of the covering fraction of $C_{F,X} > 0.85$ at the 90% significance level (see SI). The fast wind model is able to simultaneously take into account both the broad absorption at E ~ 7−10 keV and the excess of flux at E = 15−25 keV (see panel c of Extended Data Figure 1).

We note that a more physical model of a high column wind should include both Compton scattering and emission (for the latter see SI). A few of such models[33,34] have been recently reported in the literature, however they are not publicly available for use in

XSPEC and they require a relative fine-tuning of the parameters. Moreover, the wind parameters are well approximated with XSTAR absorption tables[33,34], supporting our conclusions. The effects of Compton scattering should be marginal, introducing a continuum break at an energy beyond the observed *Suzaku* bandpass of E ~ 50 keV and contribute to the broadening of the lines of less than ~ 0.3 keV.

**Slow wind model.** Considering the slow wind model, in which the feature at E ~ 9keV is identified with an ionized Fe K absorption edge, the data require a lower XSTAR turbulent velocity width of 500 km s$^{-1}$. This provides an overall good fit of the data, with $\chi^2/\nu$ = 1439.3/1391. The relative column density, ionization and outflow velocity are $N_H$ = $(5.5^{+0.6}_{-0.7}) \times 10^{24}$ cm$^{-2}$, log$\xi$ = $3.77^{+0.08}_{-0.07}$ erg s$^{-1}$ cm and $v_{out}$ = $0.0258^{+0.0109}_{-0.0128}$ $c$, respectively. The lower limit of the covering fraction is $C_F$ > 0.93 at the 90% significance level. Even though it provides an overall sufficient representation of the data, this fit is statistically much worse than the previous case of a fast wind (see panel d of Extended Data Figure 1). In fact, the $\chi^2$ difference for the same number of degrees of freedom with respect to the fast wind model is $\Delta\chi^2$ = 28.9. In particular, this slow wind model does not provide a good fit for the absorption trough at E ~ 7−10 keV.

**Relativistic reflection model.** We use the most accurate relativistic reflection code available in the literature, the "relxilllp" model[35]. This model considers the lamp post geometry in which the compact X-ray source is located at a certain height along the rotation axis of the black hole in units of gravitational radii $r_g$ = $GM_{BH}/c^2$. The reflection fraction and emissivity index are self-consistently calculated by the model depending on the source height and black hole spin. We consider a typical outer disk radius of $r_{out}$ =

400 $r_g$ and the inner radius $r_{in}$ is linked to the innermost stable circular orbit (ISCO) for a given black hole spin value. We consider standard Solar abundances.

The free parameters are the height of the illuminating source $h$, the disk inclination $i$, the ionization parameter $\log\xi$, the normalization and the black hole spin $a$. This model provides a fit statistics of $\chi^2/\nu = 1413.2/1390$. The best-fit parameters suggest a source height of just $h = 2.2^{+1.3}_{-0.6}$ $r_g$ ($r_g = GM_{BH}/c^2$) and a rapidly spinning black hole with $a > 0.85$. The material is mildly ionized, with $\log\xi = 3.15^{+0.15}_{-0.09}$ erg s$^{-1}$ cm, and the disk inclination is estimated to be $i = 52^{+4}_{-6}$ degrees. Letting the reflection parameter free to vary we find a very high best-fit value of $R \sim 6$. The relativistic reflection model provides a relatively good representation of the data up to E $\sim$ 10 keV but it is not able to model the excess of flux at E = 15−25 keV (see panel e of Extended Data Figure 1).

**Time-resolved spectral analysis.** The source lightcurve shows a factor of 10 variability in the 4−10 keV countrate over an interval of ~3 days between the minimum and maximum values (Extended Data Figure 2). This variability must be driven by the change of some parameters, such as the continuum, absorption or reflection. We performed a time resolved spectral analysis by splitting the observation at 377 ks after the beginning (Extended Data Figure 2). This allows to have the same signal-to-noise in the E = 4−10 keV band for the low-flux (first) and high-flux (second) intervals. The source shows a much smaller variability of a factor of ~1.5 in the PIN data at E = 15−25 keV. The broad-band spectra extracted in the low-flux and high-flux intervals are shown in panel a of Extended Data Figure 3.

We performed a combined fit of the low-flux and high-flux intervals using alternatively the fast wind model and the relativistic reflection model and compared the values of the parameters to determine what is the main driver of the variability in these two cases. Regarding the fit with the fast wind model, we test three different variability cases: 1) constant fully covering absorption with variable power-law continuum; 2) constant power-law continuum with variable partial covering absorption; 3) constant power-law continuum and variable fully covering absorption. The cases 1) and 3) provide the best representations of the data with comparable statistics of $\chi^2/\nu = 1757.9/1692$ and $\chi^2/\nu = 1757.3/1693$, respectively. In the case 1) of constant absorber and variable continuum we find that the main change in the fit parameters may be due to an increase in the normalization of the continuum with $\Gamma \sim 2$. The continuum slope is consistent in the two intervals. The relative ionizing luminosity between 1−1000 Ryd (1 Ryd = 13.6 eV) would be increasing from $\log L_{ion} \sim 45.2$ erg s$^{-1}$ to $\log L_{ion} \sim 45.7$ erg s$^{-1}$. Given that the column density and velocity of the absorber are constant, this increase in luminosity by a factor of ~3.4 should have caused an increase in ionization by the same factor, fully ionizing the gas, to the point that it would have not been detected in the second interval. This consideration alone strongly disfavor the continuum variability case. In the case 3) of a constant continuum and variable absorber we find that the main driver of the flux variability is a 40% decrease in column density from $N_H \sim 8.1 \times 10^{24}$ cm$^{-2}$ to $N_H \sim 4.9 \times 10^{24}$ cm$^{-2}$ between the two intervals (Extended Data Table 2). Moreover, the fact that the observed change in flux of a factor of ~10 is much larger at E= 4−10 keV compared to only a factor of ~1.5 at E > 15 keV also points to the fact that the main variability is due to a change in absorbing column density rather than continuum emission with constant slope (see panel b of Extended Data Figure 3).

Regarding the fit with the relativistic reflection model, we tested two different variability cases: 1) constant reflection with variable power-law continuum and 2) constant power-law continuum with variable reflection (i.e., source height and ionization). The case 1) would indicate a constant and very small source height of $h < 2$ $r_g$. Instead, the continuum slope would be required to change from the unrealistically high value of $\Gamma \sim 2.7$ to $\Gamma \sim 2.2$ between the two intervals. Overall, this case does not provide a good fit to the data, with a poor statistics of $\chi^2/\nu = 1878.7/1692$. The case 2) provides a better fit of $\chi^2/\nu = 1766.4/1693$. The fit with the relativistic reflection model provides a relatively good representation of the data up to $E \sim 10$ keV but not at higher energies (see panel b of Extended Data Figure 3). In this case, the direct continuum is required to be constant between the two intervals and the main driver of the variability would be the height of the X-ray source, decreasing from $h = 5.4^{+54.4}_{-1.2}$ $r_g$ down to $h < 3.2$ $r_g$ from the low-flux to the high-flux case, respectively (Extended Data Table 3). This corresponds to an increase in reflection fraction from ~3 in the low-flux interval up to the maximum value allowed of ~10 for the high flux case. Even though this fit might provide a relatively good representation of the data, it clearly contradicts the variability expected from the associated relativistic light bending case[25], which predicts a very specific variability pattern of higher/lower reflection fractions (lower/higher X-ray source heights) for lower/higher fluxes.

*Suzaku* **XIS background and systematics check.** We investigated the existence of the broad absorption at $E \sim 9$ keV also performing independent fits of the *Suzaku* XIS0, XIS1 and XIS3 detectors. In all cases the parameters are consistent at the 90% level and in particular the broad absorption is independently detected at the 3σ level in the XIS0 and XIS1 and at 4σ level for the XIS3 (Figure 1). The XIS cameras have an instrumental background emission line from Ni Kα at the observed energy of 7.47 keV. The line intensity is slightly dependent on the location on the detector. Therefore, an incorrect background selection might possibly induce spurious absorption/emission lines in the background subtracted spectrum. We performed several tests in order to exclude this possibility. First, we note that the observed centroid energy of the broad line of $E = 8.26^{+0.64}_{-0.34}$ keV is not consistent with the energy of the instrumental line. Second, the background emission line is unresolved, with a width of $\sigma_E < 100$ eV. This is not consistent with the large width of the absorption feature of $\sigma_E \sim 1$ keV. Third, we performed fits using background regions extracted from one, two and three circles of 2.5'' size away from the source and annuli with inner/outer radii of 3''/5'' and 4''/5'' centered on the source. Fourth, we performed an additional consistency check considering a smaller circular source extraction region of 1.5'' and an annular background region of the same area with inner/outer radii of 3''/3.4''. This allows us to derive a cleaner extraction region, with a marginal loss of 20% of the source counts in the $E = 4-10$ keV, and to drastically minimize the background, cutting down its countrate by a factor of 2.5. In all the cases we find that the best-fit parameters, and in particular those of the broad absorption feature, are always consistent within the 1σ errors. Considering only the XIS data and fitting both the fast and slow wind models and the relativistic reflection model we obtain parameters which are consistent within the 1σ errors with those derived using the combined XIS and PIN data. This indicates that these models depend only weakly on the E = 15−25 keV PIN data. The fast wind ($\chi^2/\nu$ = 1372.1/1365) and the relativistic reflection ($\chi^2/\nu$ = 1369.2/1364) models provide statistically comparable fits of the XIS data alone. Instead, the fit using the slow wind model is statistically worse ($\chi^2/\nu$ = 1404.6/1365).

**Modeling of the *Herschel* OH spectrum**. The profile of the OH 119 doublet shows[6] a strong asymmetry at systemic velocities and an obvious blue wing extending well beyond ≈1,000 km s$^{-1}$ (see Figure 2). This profile is modeled[7] in spherical symmetry as an extended 300 pc scale outflow. The Spitzer IRS low-resolution spectra, IRAS fluxes, and *Herschel* spectrum around the OH doublet are used to estimate the underlying continuum. The covering factor of the outflow is estimated to be $C_{F,OH}$ = 0.20 ± 0.05. The sum of the covering factors of the outflow and systemic components is close to unity, consistent with the fraction of gas affected/unaffected by the outflow. The outflow velocity is also well determined, with an average value of $v_{out,OH}$ = 1,000 ± 200 km s$^{-1}$. The absorption is seen in front of the far-IR continuum, which is well modeled with a dust radius of $R_{dust}$ = 250 pc. The distance of the OH absorber is estimated at $r$ = 300 pc in radius (or 600 pc in diameter). This value is consistent with previous results[7] derived using the same model for the more detailed spectrum of Mrk 231, indicating that the outflowing OH is mostly linked to the source of far-IR continuum.

The Hydrogen number density is derived from the equation $n_H = N_{OH}/(X_{OH} \times \Delta r)$, where $N_{OH}$ is the OH column density, $X_{OH}$ is the OH abundance relative to H and $\Delta r$ is the width of the region sampled by the outflowing OH. The estimated value of the OH column density from the fit of the line profile is $N_{OH}$ = 3×10$^{16}$ cm$^{-2}$. This value is conservatively low but enough to account for the optically thick OH119 absorption line.

The standard value[7] of the OH abundance relative to H in active regions where OH is highly abundant is $X_{OH} = 2 \times 10^{-6}$. The width of the outflowing OH region of $\Delta r = 50$pc is derived from the relation $R_{out} = 1.2\, R_{dust}$, which is based on more accurate models[7] for the similar OH profile in Mrk 231. Therefore, we estimate a Hydrogen number density of $n_H = 100$ cm$^{-3}$. Exploring the parameter space of the model, we find an uncertainty of a factor of two on this value (see SI).

31. Kallman, T., Bautista, M. Photoionization and High-Density Gas. *Astrophys. J. Suppl. Ser.* **133**, 221−253 (2001)

32. Fukumura, K. *et al.* Stratified Magnetically Driven Accretion-disk Winds and Their Relations to Jets. *Astrophys. J.* **780**, 120 (2014)

33. Sim, S. A., Miller, L., Long, K. S., Turner, T. J., Reeves, J. N. Multidimensional modelling of X-ray spectra for AGN accretion disc outflows – II. *Mon. Not. R. Astron. Soc.* **404,** 1369-1384 (2010)

34. Hagino, K., Odaka, H., Done, C., *et al.* The origin of ultra-fast outflows in AGN: Monte-Carlo simulations of the wind in PDS 456. *Mon. Not. R. Astron. Soc.* **446**, 663-676 (2015)

35. García, J. *et al.* Improved Reflection Models of Black Hole Accretion Disks: Treating the Angular Distribution of X-Rays. *Astrophys. J.* **782**, 76 (2014)

**Extended Data Figure 1. Broad-band *Suzaku* spectrum in the E=0.5-30 keV band. a**, the time-averaged *Suzaku* XIS03 (solid black), XIS1 (dotted red) and PIN (solid green) spectra binned to 10σ, 5σ and 3σ, respectively. The data to model residuals in units of sigma with respect to the absorbed power-law model, the fast wind model, the slow wind model and the relativistic reflection model are shown in panel **b**, **c**, **d** and **e**, respectively. The energy is in the rest-frame and errors are at the 1σ level.

**Extended Data Figure 2. Background subtracted *Suzaku* XIS03 light curve in the E=4−10 keV band.** The data are binned to the *Suzaku* orbital period of 5760 seconds. The vertical line indicates the time at which the observation is split in two parts for the time-resolved spectral analysis. The effective on-source exposure time is 250 ks. The gaps in the light curve indicate periods in which the satellite could not point to the source. Therefore, the total temporal coverage of the observation is longer, about 500 ks. Errors are at the 1σ level.

**Extended Data Figure 3. Time-resolved *Suzaku* spectral analysis in the E=0.5−30 keV band. a,** *Suzaku* XIS03 and PIN spectra extracted during the low-flux (green) and high-flux (blue) intervals. The XIS03 and PIN data are rebinned to 10σ and 5σ, respectively. The data to model ratios with respect to the fast wind model and relativistic reflection model are reported in panel **b** and **c**, respectively. Errors are at the 1σ level.

**Extended Data Table 1. Best-fit values of the fast wind model for the time-averaged spectrum.** Errors are at the 1σ level.

**Extended Data Table 2. Best-fit values of the fast wind model for the time-resolved spectral analysis.** *Notes.* Errors are at the 90% level.

**Extended Data Table 3. Best-fit values of the relativistic reflection model for the time-resolved spectral analysis.** *Notes.* The parameter *a* and *R* refer to the black hole

spin and reflection fraction, respectively. Errors are at the 90% level.

**Extended Data Table 4. Parameters of outflows in other quasars and ULIRGs collected from the literature.** *Notes.* The reported velocity refers to the average outflow velocity. The parameters $\dot{P}$ and $L_{AGN}/c$ indicate the momentum flux of the outflow and AGN radiation, respectively. Errors are at the 1σ level. The last column indicates the relevant references. The parameters of the inner X-ray wind and OH molecular outflow in IRAS F11119+3257 indicated by # refer to this work.

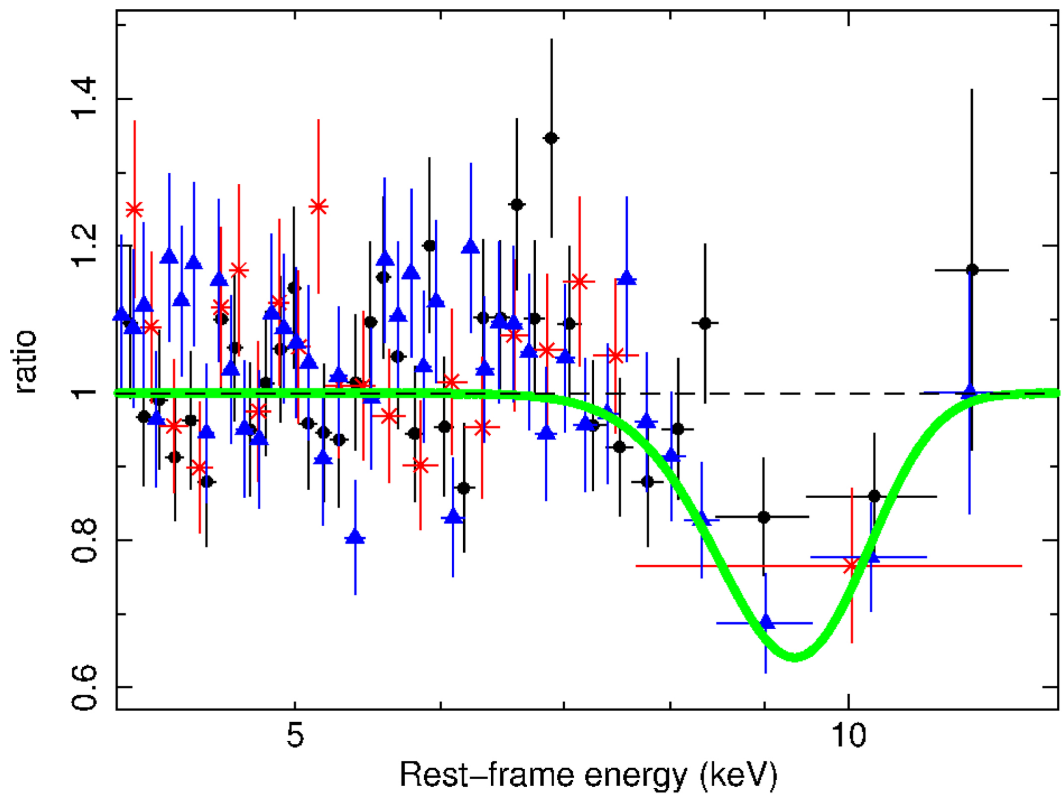

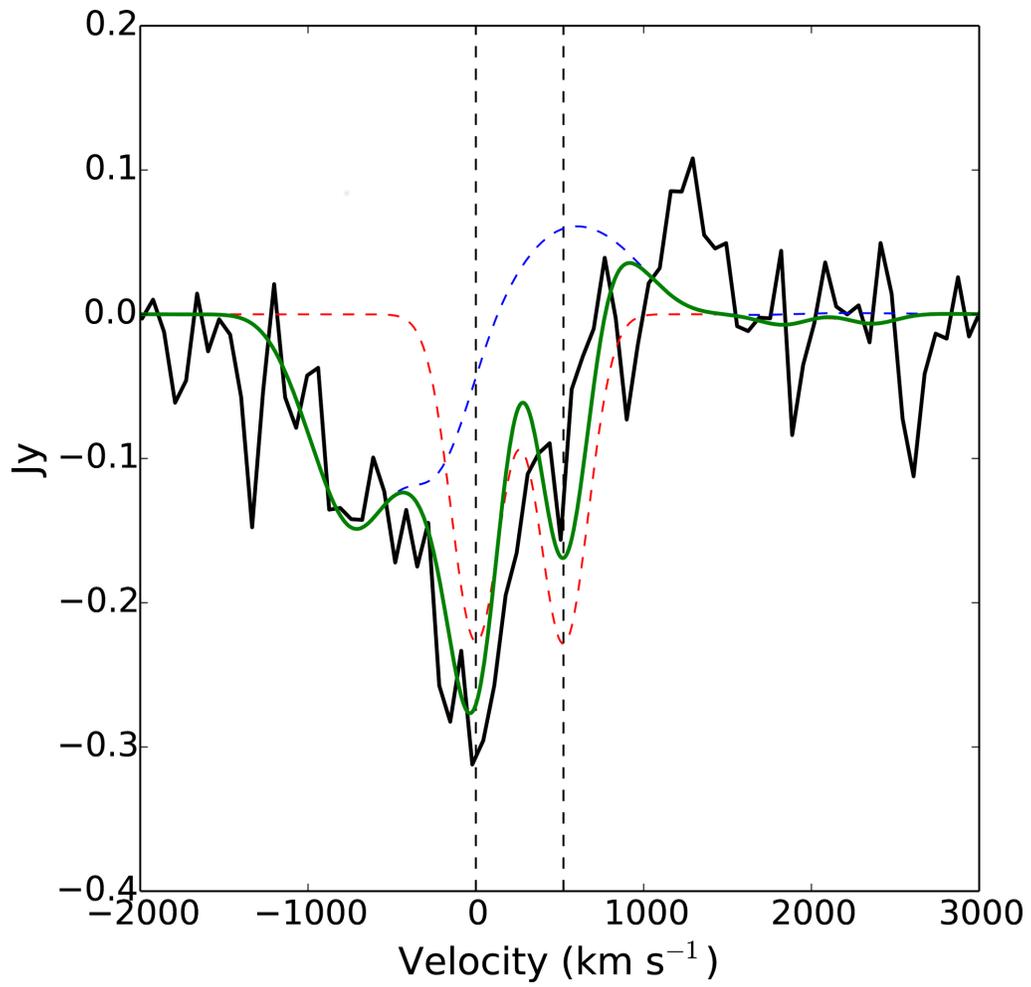

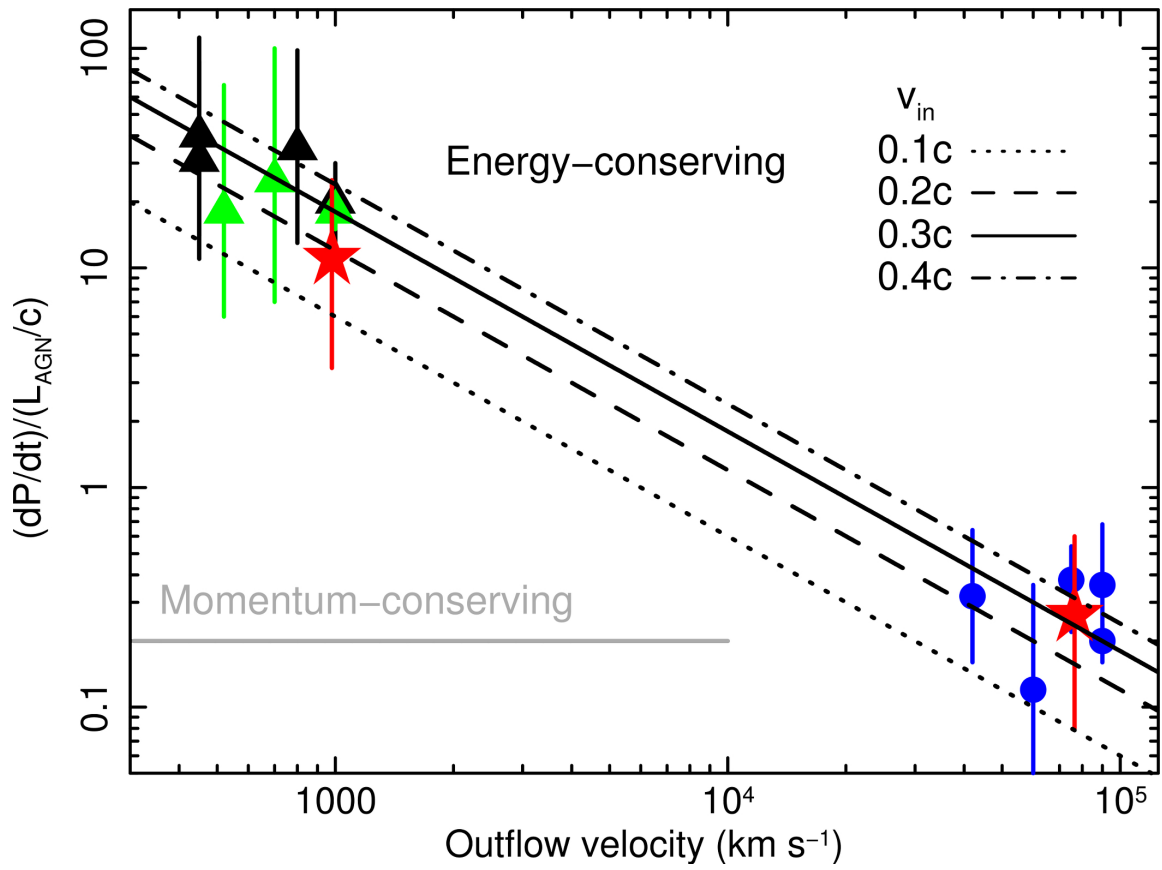

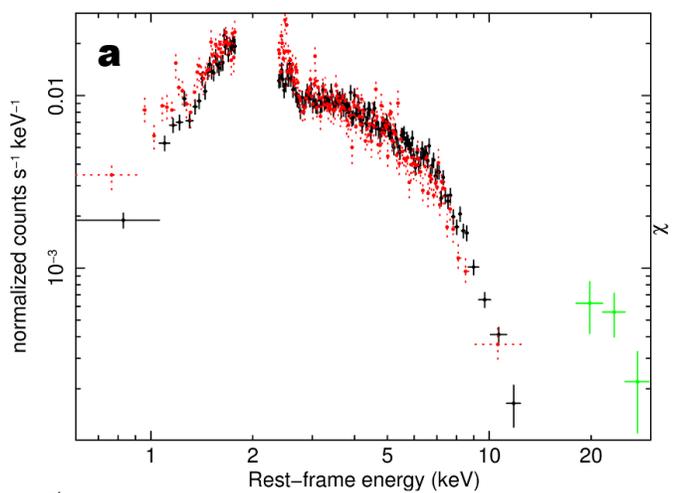
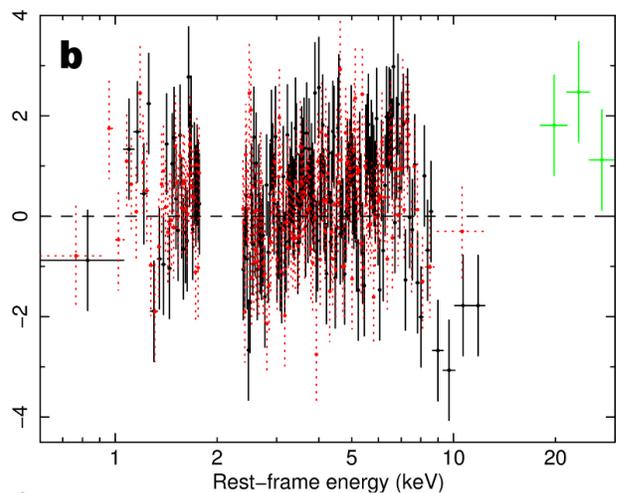
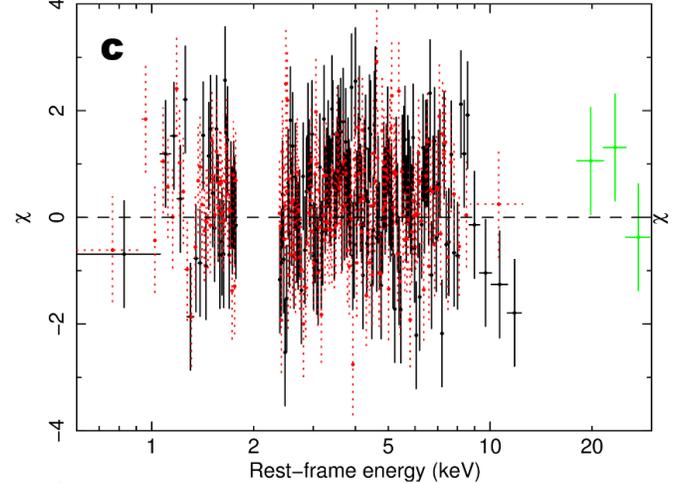
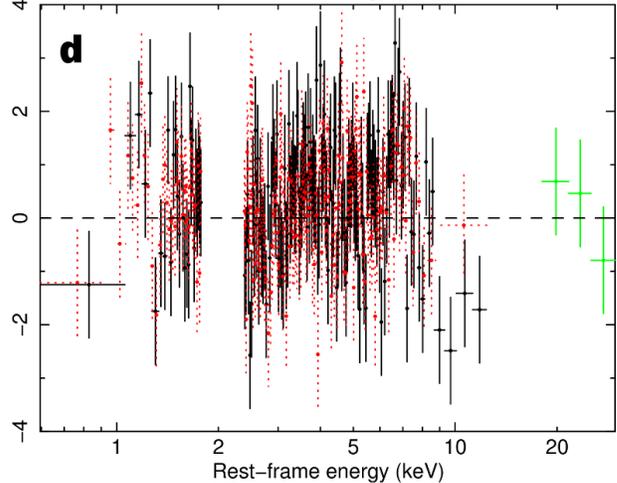
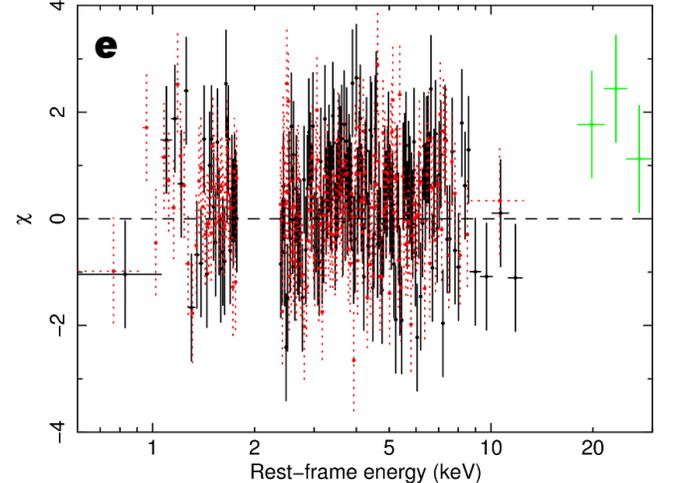

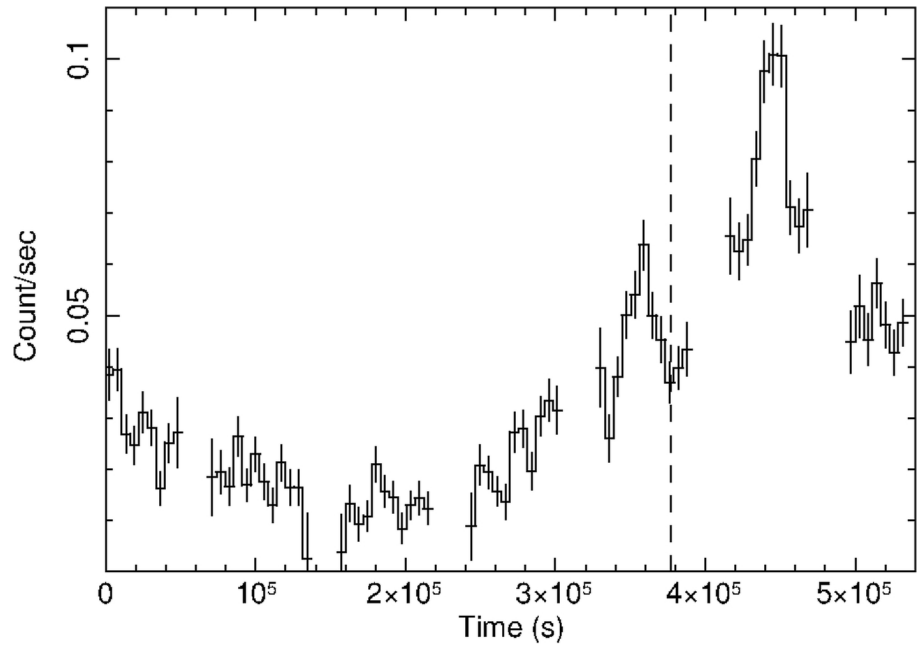

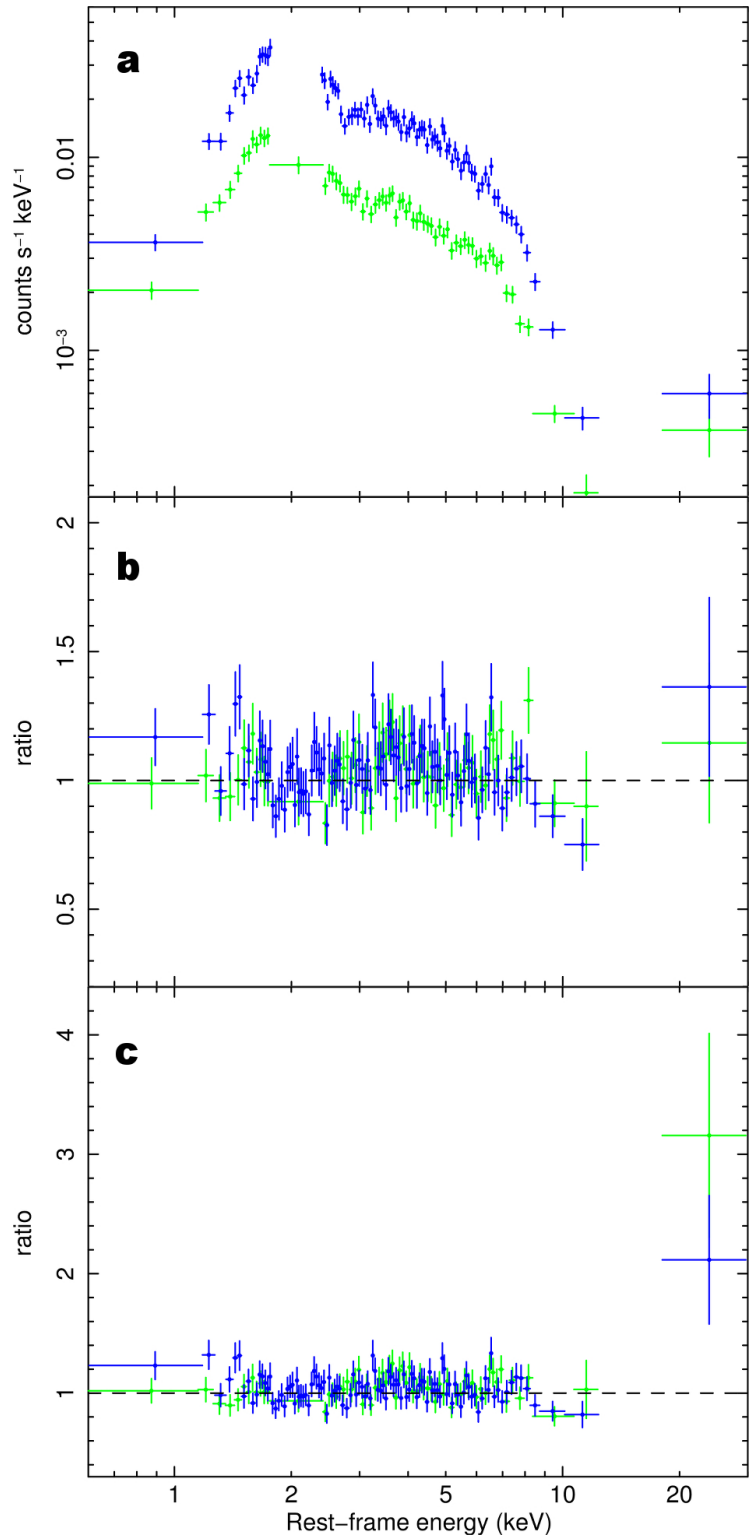

| Absorbed Power-law | | |
|---|---|---|
| $\Gamma$ | $N_H$ ($10^{22}$ cm$^{-2}$) | $\chi^2/\nu$ |
| 2.04±0.03 | 2.07±0.08 | 1410.4/1391 |
| Photoionized absorption | | | |
|---|---|---|---|
| $\log\xi$ (erg s$^{-1}$ cm) | $N_H$ ($10^{24}$ cm$^{-2}$) | $v_{out}$ (c) | $C_F$ |
| $4.11^{+0.09}_{-0.04}$ | $6.4^{+0.8}_{-1.3}$ | $0.255 \pm 0.011$ | >0.85 |

| | Low-Flux | High-Flux |
|---|---|---|
| Absorbed power-law | | |
| $\Gamma$ | $2.05 \pm 0.05$ | |
| $N_H$ ($10^{22}$ cm$^{-2}$) | $2.06 \pm 0.12$ | |
| Photoionized absorption | | |
| $\log\xi$ (erg s$^{-1}$ cm) | $4.08^{+0.08}_{-0.03}$ | $4.13^{+0.20}_{-0.08}$ |
| $N_H$ ($10^{24}$ cm$^{-2}$) | $8.1^{+0.6}_{-1.1}$ | $4.9^{+1.3}_{-1.9}$ |
| $v_{out}$ (c) | $0.243^{+0.020}_{-0.021}$ | $0.268^{+0.031}_{-0.034}$ |
| Best-fit statistics | | |
| $\chi^2/\nu$ | 1757.3/1692 | |

|  | Low-Flux | High-Flux |
|---|---|---|
| **Absorbed power-law** | | |
| $\Gamma$ | $2.01^{+0.08}_{-0.04}$ | |
| $N_H$ ($10^{22}$ cm$^{-2}$) | $2.24^{+0.15}_{-0.10}$ | |
| **Relativistic reflection** | | |
| $a$ | $> 0.93$ | |
| $h$ ($r_g$) | $5.4^{+54.4}_{-1.2}$ | $< 3.2$ |
| $i$ (degrees) | $44^{+5}_{-7}$ | |
| $\log\xi$ (erg s$^{-1}$ cm) | $3.05^{+0.13}_{-0.07}$ | $3.32^{+0.13}_{-0.09}$ |
| $R$ | $\simeq 3$ | $\simeq 10$ |
| **Best-fit statistics** | | |
| $\chi^2/\nu$ | 1766.4/1693 | |

| Name | $\log L_{AGN}$ (erg s$^{-1}$) | $v_{out}$ (km s$^{-1}$) | $\dot{P}/(L_{AGN}/c)$ | Refs |
|---|---|---|---|---|
| **Quasars (X-rays)** | | | | |
| IRAS F11119+3257 | 46.2 | 75000 | $1.3^{+1.7}_{-0.9}$ | # |
| APM 08279+5255 | 47.0 | 60000 | $0.6^{+1.2}_{-0.4}$ | 15 |
| PG 1211+143 | 45.7 | 42000 | $1.6^{+1.6}_{-0.8}$ | 16 |
| HS 1700+6416 | 47.0 | 90000 | $1 \pm 0.2$ | 17 |
| HS 0810+2554 | 46.0 | 90000 | $1.8^{+1.6}_{-1.1}$ | 18 |
| PDS 456 | 47.0 | 75000 | $1.9 \pm 0.8$ | 19 |
| **ULIRGs (OH)** | | | | |
| IRAS F11119+3257 | 46.2 | 1000 | $11.0^{+14.1}_{-7.5}$ | # |
| IRAS 08572+3915 | 45.7 | 700 | $25^{+75}_{-18}$ | 5 |
| IRAS 13120−5453 | 44.9 | 520 | $18^{+50}_{-12}$ | 5 |
| Mrk 231 | 46.0 | 1000 | $18 \pm 3$ | 7 |
| **ULIRGs (CO)** | | | | |
| IRAS F08572+3915 | 45.7 | 800 | $35^{+63}_{-22}$ | 8 |
| IRAS F10565+2448 | 44.8 | 450 | $40^{+72}_{-26}$ | 8 |
| IRAS 23365+3604 | 44.7 | 450 | $31^{+56}_{-20}$ | 8 |
| Mrk 231 | 46.0 | 1000 | $20 \pm 10$ | 4 |

**Supplementary Information**

**1. Phenomenological modeling of the *Suzaku* spectra**

The baseline continuum model required to characterize the underlying continuum consists of a power-law absorbed by a neutral column of gas and a scattered (unabsorbed) power-law soft X-ray continuum. We obtain a good fit ($\chi^2/\nu$ = 1464.6/1394) with parameters $N_H = (2.15 \pm 0.08) \times 10^{22}$ cm$^{-2}$, $\Gamma = 2.00 \pm 0.03$ and a scattering fraction of ~7%.

The data to model residuals in units of sigma with respect to the baseline continuum model show residuals in absorption and emission at rest-frame energies of E ~ 7−12 keV and E ~ 20−30 keV, respectively (panel b of Extended Data Figure 1). In particular, the Fe K band residuals are very clear if considering a ratio of the spectrum zoomed between E = 4−12 keV with respect to the absorbed power-law continuum (Figure 1).

The inclusion of a broad Gaussian absorption line in the broad-band spectrum provides a very high improvement of the fit, with $\Delta\chi^2/\Delta\nu$ = 52.9/3, corresponding to a statistical confidence level of 6.5σ. The final best-fit statistics is $\chi^2/\nu$ = 1411.7/1391. The parameters of the absorption line are $E = 9.82^{+0.64}_{-0.34}$ keV, $\sigma_E = 1.67^{+1.00}_{-0.44}$ keV and EW = $-1.31^{+0.40}_{-0.31}$ keV.

A possible alternative phenomenological modeling of the broad absorption trough could be provided by an absorption edge. The inclusion of an absorption edge to the baseline continuum model provides a best-fit of $\chi^2/\nu$ = 1423.3/1392. The edge energy and optical depth are $E = 8.60 \pm 0.07$ keV and $\tau = 0.51 \pm 0.10$, respectively. The broad Gaussian absorption line still provides a statistically better fit.

A factor of ~2−3 excess of emission in the PIN spectrum at E=15−25 keV is present after the inclusion of the broad Gaussian absorption line or the edge. The statistical significance of this excess is at the ~ 4σ level. We note that this is a conservative estimate, which takes into account of both the statistical and systematic uncertainties in the latest version of the PIN background (see Methods).

**2. Fast wind model**

Following the interpretation as broad absorption line, the residuals at the energy of E ~ 9 keV could be identified with blue-shifted Fe XXV Heα and/or Fe XXVI Lyα resonant absorption with a velocity of ~0.3$c$. If due to a single line, the measured width would indicate a large velocity broadening of $\sigma_v$ ~ 50,000 km s$^{-1}$. We checked that the alternative possibility of a blend of two Fe XXV and Fe XXVI absorption lines ($\sigma_v$ ~ 10,000 km s$^{-1}$) with energies fixed to the values of 6.7 keV and 6.97 keV and a common blue-shifted velocity of ~0.35c provides a poorer fit compared to a single broad line ($\chi^2/\nu$ = 1434.1/1391).

The output parameters of the XSTAR fit are the column density, ionization parameter and the observed absorber redshift $z_o$. The observed absorber redshift is related to the intrinsic absorber redshift in the source rest frame $z_a$ as $(1+z_o) = (1+z_a)(1+z_c)$, where $z_c$ is the cosmological redshift of the source. The velocity can then be determined using the relativistic Doppler formula $1+z_a = (1-\beta / 1+\beta)^{1/2}$, where $\beta = v/c$ is here required to be positive for an outflow. This ensures that the relativistic effects associated with both high red-shift sources and high-velocity outflows are correctly taken into account when inferring absorber outflow velocities relative to the source rest-frame.

We estimate the partial covering parameter of the absorber considering a double power-law continuum with the same photon index and free normalizations. The first represents the unabsorbed continuum. The second instead is absorbed by the XSTAR table. The two normalizations are $N_U$ and $N_A$, respectively. Then, the covering fraction can be easily calculated as $C_{F,X} = N_A / (N_U + N_A)$.

Even if it is not clear from the spectrum (Figure 1), we checked the possibility of detecting emission from the wind. We calculated an XSTAR emission table with the same parameters as for the absorption table (see Methods). The exact characteristics of the associated P-Cygni profile depend on the geometry of the wind. We note that the P-Cygni profile from an accretion disk wind may deviate from the spherical symmetry assumption. In fact, if the wind is observed relatively close to the accretion disk, then the receding part may be occulted by the disk itself and we would expect a net blue-shifted velocity for the emission as well. Moreover, for a large covering fraction, we would also expect a width of the emission component comparable to the outflow velocity of the wind, modulo a projection factor. We estimate the possible P-Cygni profile linking the ionization parameter and column density between the emission and absorption tables. The emission table was convolved with a Gaussian broadening profile. We obtain only a marginal (~99%) fit improvement of $\Delta\chi^2 = 12.3$ for three additional degrees of freedom. The resultant emission is broad, with $\sigma_E \sim 0.5 keV$, and it has a velocity shift comparable to that of the absorber. The covering fraction is consistent between the emission and absorption, with a value $C_{F,X} \sim 1$. Given the marginal requirement of the emission table and the fact that the parameters of the absorber are consistent at the 1σ level with or without its inclusion, we do not discuss it further. However, it provides additional support for the presence of a wide-angle wind with a high covering fraction of $C_{F,X} \sim 1$.

## 3. Mid-infrared emission line diagnostics

Another important test to further distinguish between the relativistic reflection and the fast wind cases is the comparison of the relative luminosities with that expected from line diagnostics. Mid-infrared emission line diagnostics can be used to estimate the expected AGN hard X-ray luminosity[36,37]. Considering the luminosities[38,39] of the Ne III (15.51 μm), Ne V (14.32 μm) and O IV (25.89 μm) we estimate an expected hard X-ray luminosity of $\log L = 45.1$ erg s$^{-1}$ in the E = 14−195 keV energy band. Extrapolating the absorption corrected luminosity of the fast wind model we obtain exactly this value. Instead, the relativistic reflection model underestimates this luminosity by a factor larger than 10.

## 4. Fast wind characteristics

### 4.1. Fast wind location

In the following we use the parameters derived from the time-averaged fast wind model (Extended Data Table 1). The ionization parameter is defined[40] as $\xi = L_{ion}/nr^2$, where $n$ is the density of the material, $r$ is the distance to the source and $L_{ion}$ is the ionizing luminosity. The absorption corrected ionizing luminosity from the wind model is $\log L_{ion} = 45.5$ erg s$^{-1}$. A lower limit on the location of the absorber can be derived from the radius at which the observed velocity corresponds to the escape velocity, $r = 2G M_{BH}/v_{out,X}^2 \sim 7.91 \times 10^{13}$ cm. Converting in units of Schwarzschild radii, the wind is observed at a distance of $r \sim 15$ $r_s$ from the central black hole.

The time-resolved spectral analysis indicates a column density decrease from $N_l = 8.1 \times 10^{24}$ cm$^{-2}$ to $N_h = 4.9 \times 10^{24}$ cm$^{-2}$ in ~3 days between the low and high flux intervals, respectively (Extended Data Table 2). The ionization parameter and outflow velocity do not vary significantly. Combining the definition of the ionization parameter $\xi = L_{ion}/nr^2$ and the column density $N_H = n\Delta r \sim nr$ for this compact absorber we find the relation $N_l\, r_l \sim N_h\, r_h$. This indicates that a decrease in column density without an increase in ionization parameter and luminosity would correspond to a shift in location $\Delta r = r_h - r_l$ between the two intervals. Considering the average outflow velocity of $v_{avg} \sim 0.25c$ and the variability occurring on a time interval of $\Delta t \sim 3$ days, we derive $\Delta r \sim v_{avg} \times \Delta t$. The value $\Delta r$ is an upper limit given that the transverse or expansion velocity might be lower than the outflow velocity. Combining these equations we obtain a relation between the initial location and the other observables: $r_l = v_{avg}\, \Delta t\, N_h/(N_l - N_h)$. Substituting the relative values, we can estimate an upper limit of the absorber location of $r_l < 5 \times 10^{15}$ cm, corresponding to less than 0.002 pc (or 900 $r_s$) from the black hole.

A less stringent upper limit on the location of this compact absorber can also be estimated from the ionization parameter $r < L_{ion}/\xi N_H \sim 3.7 \times 10^{16}$ cm. Even assuming this upper limit, the wind is found within a distance of less than 0.01 pc from the black hole. Indeed, this indicates that this compact absorber is directly identifiable with an accretion disk wind.

**4.2. Fast wind energetics**

The mass outflow rate of the fast wind can be estimated considering the equation[41] $\dot{M}_{out,X} = 4\pi\, \mu\, m_p\, r\, N_{H,X}\, C_{F,X}\, v_{out,X}$, where $\mu = 1.4$ is the mean atomic mass per proton, $m_p$ is the proton mass, and $C_{F,X}$ is the wind covering fraction. In spherical symmetry this latter value corresponds to the solid angle subtended by the wind of $\Omega = 4\pi C_{F,X}$. We conservatively consider the distance of $r = 15\ r_s$ from the black hole. Substituting the relative best-fit values (Extended Data Table 1) we obtain an estimate of $\dot{M}_{out,X} = 1.5\ M_\odot\ yr^{-1}$. The momentum flux (or force) and mechanical (or kinetic) power of the wind can then be estimated as $\dot{P}_{out,X} = \dot{M}_{out,X}\ v_{out,X}$ and $\dot{E}_{K,X} = (1/2)\dot{M}_{out,X}\ v_{out,X}^2$, respectively. Given the very high luminosity of this AGN and the fact that it is accreting at about its Eddington limit, it is very likely that the wind is radiation driven[21]. For a highly ionized wind driven by radiation pressure we have the relation $\dot{P}_{out,X} = C_F\ \tau\ \dot{P}_{rad}$, where $C_F$, $\tau$ and $\log\dot{P}_{rad} = L_{AGN}/c = 35.7$ dyne are the wind covering fraction, optical depth and the momentum flux of the AGN radiation, respectively. Given the wind covering fraction of $C_F \sim 0.85$ and the optical depth of $\tau \sim 3.5$, we can estimate a maximum value of $\dot{P}_{out,X} \sim 3\ \dot{P}_{rad}$. This is consistent with multiple scatterings in a high column wind[22,23] and which covers a large solid angle, so that multiple photon scatterings can occur within the wind. Therefore, considering the uncertainty on the black hole mass[20] and the limit on the wind momentum, the momentum rate and mechanical power of the fast wind are $\log\dot{P}_{out,X} = 35.8^{+0.4}_{-0.5}$ dyne and $\log\dot{E}_{K,X} = 45.4^{+0.4}_{-0.5}$ erg s$^{-1}$, respectively. The ratio of the momentum flux of the initial fast wind and that of the AGN radiation is $\dot{P}_{out,X} = 1.3^{+1.7}_{-0.9}\ \dot{P}_{rad}$. The wind power corresponds to ~15% of the AGN luminosity. This is significantly higher than the minimum of ~0.5% required for AGN feedback[24].

5. Energetics of the large-scale molecular outflow

The mass outflow rate of the OH outflow can be estimated considering the equation[7]

$\dot{M}_{out,OH} = 800 \times (C_{F,OH}/0.2) \times (n_H/100\ cm^{-3}) \times (r/300\ pc)^2 \times (v_{out,OH}/1{,}000\ km\ s^{-1})\ M_\odot\ yr^{-1}$. Here, $C_{F,OH}$ is the wind covering fraction, $n_H$ is the Hydrogen number density, $r$ is the radius of the absorber and $v_{out,OH}$ is the outflow velocity. This equation assumes a spherical, but clumpy, geometry. This geometry is consistent with a wide-angle outflow, as inferred from the strong asymmetry between the two doublet components indicating roughly similar gas masses flowing on the near and back sides of the far-IR source.

We find a best-fit value of the Hydrogen number density of $n_H \sim 100$ cm$^{-3}$ (see Methods). In principle, solutions with thickness of the absorbing shell approaching the size of the far-IR source of up to $\Delta r \sim 100$pc can match the OH119 spectrum, suggesting Hydrogen densities as low as $n_H \sim 50$ cm$^{-3}$. However, there are both observational and theoretical arguments favoring thinner shells and consequently higher densities. First, translucent clouds have typical Hydrogen number densities of $n_H \sim 50-100$ cm$^{-3}$. However, if these regions are compressed by the passage of a shock we can expect $n_H$ to be significantly higher than 50 cm$^{-3}$. Second, the circumnuclear regions where these outflows emerge have significantly higher average densities of $n_H \geq 1,000$ cm$^{-3}$. Third, the OH excitation observed in the similar outflow[7] in Mrk 231 strongly favors the thin-shell approximation with $\Delta r/r < 1$. Fourth, the thin-shell approximation is consistent with the prediction of theoretical models[9,10,11] discussing the origin and structure of large-scale molecular outflows. Therefore, we estimate a minimum value for the mass outflow rate of 250 M$_\odot$ yr$^{-1}$. Conversely, values higher than 2,000 M$_\odot$ yr$^{-1}$ are disfavored from the model because they would indicate too thin shells with a width of just a few parsecs.

Therefore, taking into account the uncertainties of the parameters in the equation (see Methods), the mass outflow rate can be confidently estimated within a factor of three. Substituting the relative values we derive $\dot{M}_{out,OH} = 800^{+1,200}_{-550}$ M$_\odot$ yr$^{-1}$. The resultant momentum flux of the OH wind is $\log \dot{P}_{out,OH} = 36.7 \pm 0.5$ dyne. This is significantly larger than the momentum rate of the AGN radiation, $\dot{P}_{out,OH} = 11.0^{+14.1}_{-7.5} \dot{P}_{rad}$. The mechanical power of the molecular outflow is $\log \dot{E}_{K,OH} = 44.4 \pm 0.5$ erg s$^{-1}$. This corresponds to ~2% of the AGN luminosity.

Relatively high mass outflow rates of ~1,000 M$_\odot$ yr$^{-1}$ and consequently short depletion timescales of ~10$^7$ years required to remove a fraction of the molecular gas from the inner regions of the galaxy have been reported in several ULIRGs hosting powerful molecular outflows in both OH[5,7] and CO[4,8] observations. However, this estimate of the depletion timescale is only a lower limit, given that the observed molecular outflows are likely to be non-steady, as recently suggested[7] for Mrk 231. Therefore, the reported mass outflow rate is a local estimate at R ~ 300 pc, which is valid for timescales less than ~ R/v ~ 10$^5$ years. Moreover, if we consider a supermassive black hole with M$_{BH}$ ~ 10$^7$ M$_\odot$, as for IRAS F11119+3257, accreting at its Eddington rate with a typical efficiency η ~ 0.1 we can estimate an AGN lifetime of ~3×10$^7$ years. This is comparable to the estimated depletion timescale and supports the fact that the AGN is effective at removing the molecular gas during its lifetime.

**6. Linking the nuclear X-ray wind to the large-scale molecular outflow**

In Figure 3 we do not include the ultrafast outflows detected in lower luminosity Seyferts[13,14]. However, we note that the momentum flux of these winds is also comparable to that of the AGN radiation. Thus, they will be located in the same parameter space as the accretion disk winds in quasars shown in the right-hand side of Figure 3. The molecular outflows shown in Figure 3 are selected to have average velocities higher than about 500 km s$^{-1}$, ensuring that they are most likely driven primarily by the AGN instead of the starburst[6,42]. The analytical relations[9,10,11] shown in Figure 3 represent only a rough approximation of this complex AGN wind phenomenon. However, the reported energy-conserving relation is overall confirmed by recent detailed numerical simulations[12,43,44].

36. Meléndez, M. *et al.* New Indicators for AGN Power: The Correlation between [O IV] 25.89 μm and Hard X-Ray Luminosity for Nearby Seyfert Galaxies. *Astrophys. J.* **682**, 94-103 (2008)

37. Weaver, K. A. *et al.* Mid-infrared Properties of the Swift Burst Alert Telescope Active Galactic Nuclei Sample of the Local Universe. I. Emission-line Diagnostics. *Astrophys. J.* **716**, 1151-1165 (2010)

38. Veilleux, S. *et al.* Spitzer Quasar and Ulirg Evolution Study (QUEST). IV. Comparison of 1 Jy Ultraluminous Infrared Galaxies with Palomar-Green Quasars. *Astrophys. J. Suppl. Ser.* **182**, 628-666 (2009)

39. Farrah, D. *et al.* High-Resolution Mid-Infrared Spectroscopy of Ultraluminous Infrared Galaxies. *Astrophys. J.* **667**, 149-169 (2007)

40. Tarter, C. B., Tucker, W. H., Salpeter, E. E. The Interaction of X-Ray Sources with Optically Thin Environments. *Astrophys. J.* **156**, 943 (1969)

41. Crenshaw, D. M., Kraemer, S. B. Feedback from Mass Outflows in Nearby Active Galactic Nuclei. I. Ultraviolet and X-Ray Absorbers. *Astrophys. J.* **753**, 75 (2012)

42. Spoon, H.~W.~W. *et al.* Diagnostics of AGN-Driven Molecular Outflows in ULIRGs from Herschel-PACS Observations of OH at 119 μm. *Astrophys. J.* **775**, 127 (2013)

43. Costa, T., Sijacki, D., & Haehnelt, M. G. Feedback from active galactic nuclei: energy-versus momentum-driving. *Mon. Not. R. Astron. Soc.* **444,** 2355-2376 (2014)

44. Nims, J., Quataert, E., Faucher-Giguère, C.-A. Observational Signatures of Galactic Winds Powered by Active Galactic Nuclei. arXiv:1408.5141 (2014)